\documentclass[a4paper,twocolumn,prl,superscriptaddress, bibnotes,notitlepage,nofootinbib]{revtex4-2}

\usepackage{CJK}

\usepackage{graphicx}

\usepackage{dcolumn}

\usepackage{bm}
\usepackage[normalem]{ulem}
\usepackage{color}
\usepackage{amsmath}
\usepackage{amssymb}
\usepackage{lipsum}
\usepackage{bm}
\usepackage{xcolor}
\usepackage{graphicx}
\usepackage{verbatim}
\usepackage[T1]{fontenc}

\ExplSyntaxOn
\NewDocumentEnvironment{multiequation}{b}
 {
  \vantiempham:n { #1 }
 }
 {}

\seq_new:N \l__vantiempham_md_seq

\cs_new_protected:Nn \vantiempham:n
 {
  \seq_set_split:Nnn \l__vantiempham_md_seq { \\ } { #1 }
  $$
  \seq_map_function:NN \l__vantiempham_md_seq \__vantiempham_md_item:n
  $$
 }
\cs_new_protected:Nn \__vantiempham_md_item:n
 {
  \begin{minipage}{\dim_eval:n {\displaywidth/(\seq_count:N \l__vantiempham_md_seq)} }
  \begin{equation}
  \cs_set_eq:Nc \label { ltx@label }
  #1
  \vphantom{\seq_use:Nn \l__vantiempham_md_seq {}}
  \end{equation}
  \end{minipage}
 }

\ExplSyntaxOff  

\usepackage[breaklinks=true]{hyperref}

\newcommand{\bsf}[1]{\textsf{\textbf{#1}}}
\usepackage{epstopdf}

\newcommand{\beq}{\begin{equation}}
	\newcommand{\eeq}{\end{equation}}
\newcommand{\beqn}{\begin{eqnarray}}
	\newcommand{\eeqn}{\end{eqnarray}}

\newcommand{\bew}{\begin{widetext}}
	\newcommand{\ew}{\end{widetext}}

\usepackage{xcolor}
\definecolor{CMaroon}{HTML}{800000}

\begin{document}
	\title{Chirality destroys oriented active solids} 
   
\author{S. J. Kole}
\email{sjkole.acad@gmail.com}
\affiliation{Pritzker School of Molecular Engineering, University of Chicago, Chicago, IL 60637, USA
    }
 
\author{Gareth P. Alexander}
\email{G.P.Alexander@warwick.ac.uk}
\affiliation{Department of Physics, Gibbet Hill Road, University of Warwick, Coventry CV4 7AL, United Kingdom}
    
\author{Ananyo Maitra}
\email{nyomaitra07@gmail.com}
\affiliation{{Laboratoire de Physique Th\'eorique et Mod\'elisation, CNRS UMR 8089,
		CY Cergy Paris Universit\'e, F-95032 Cergy-Pontoise Cedex, France}}
\affiliation{Sorbonne Universit\'{e} and CNRS, Laboratoire Jean Perrin, F-75005, Paris, France}

\author{Sriram Ramaswamy}
\email{sriram@iisc.ac.in}
\affiliation{Centre for Condensed Matter Theory, Department of Physics, Indian Institute of Science, Bangalore 560 012, India}
\affiliation{International Centre for Theoretical Sciences, Tata Institute of Fundamental Research, Bangalore 560 089 India}

	
\begin{abstract}
We construct a hydrodynamic theory of the effect of chiral and active force densities on three-dimensional uniaxial solids. We show that in a fixed background medium active chiral force densities---that arise at lower order in gradients than odd elasticity---generically destabilise the solid state. We further show this active instability also affects \emph{isotropic} chiral solids in an anisotropic medium, a situation that is relevant for cell clusters moving through the extracellular matrix. Interestingly, the instability is eliminated in a higher-symmetry solid, known as the polar elastomer; indeed, we show that chirality does not affect the leading-order mechanics of such materials. We further demonstrate that momentum-conserving polar active solids have two chiral odd elasticities in contrast to the one discussed for two-dimensional solids. However, we again find that the higher symmetry of polar elastomers eliminates one of these with the remaining one being equivalent to the two-dimensional odd elastic force density. Our results should apply to numerous biological and condensed matter systems.
\end{abstract}
	
	\maketitle

The interplay of homogeneous energy injection and spontaneous or particle-scale spatial asymmetry in active systems~\cite{RMP,ramaswamy2010mechanics,bowick2022symmetry} leads to physics that is impossible in equilibrium.  
Chirality is an asymmetry whose interplay with time-irreversibility has been shown, in recent years, to lead to a variety of interesting behaviours \cite{maitra2025activity,alexander2025screw,fruchart2023odd, burekovic2026bulk}, ranging from phases in which the direction of ordering rotates spontaneously, to odd elasticity in states that break translation symmetry, to the transverse velocity-dependent ``odd'' mechanical response~\cite{maitra2019spontaneous,zhang2020reconfigurable,scheibner2020odd,bililign2022motile,tan2022odd, banerjee2017odd, khain2022stokes,markovich2024nonreciprocity, kole2025non}. However, with a few exceptions, these results are applicable to two-dimensional systems and, since chirality is intrinsically dimension-dependent, cannot be transplanted to three-dimensional (3D) materials. The origin and core motivation of active-matter theory is its status as the natural theoretical description of the mechanics and statistics of living materials. As life is three-dimensional and uncompromisingly chiral~\cite{zhang2024helical, prost2015active,naganathan2014active,tee2015cellular,inaki2018chiral, chin2018epithelial}, the study of the relatively neglected 3D active chiral solids is of both fundamental and practical importance.  

We examine the effect of chiral activity on three-dimensional anisotropic solids within a general hydrodynamic framework and demonstrate that, in an inert dissipative background, they are \emph{generically} destabilised. That is, they are unstable with zero activity threshold at asymptotically small wavenumbers. The wavelength of the fastest-growing mode for this instability scales inversely with chiral activity. The unstable mode has a helical displacement field which, we speculate, permanently endows the solid with an orientation field like that of a cholesteric with pitch given by this wavelength. Such structures are common in living tissues~\cite{zhang2024helical, Bouligand_LC,Bouligand_twisted,Heartstrings}. 

Interestingly, while well-known examples of generic instability of ordered states in active materials~\cite{Aditi, SJ_smec, RMP, AM_interface, Rayan} arise in systems with long-range interactions due to Stokesian hydrodynamics, the instability we discuss here arises in the opposite dry limit in which the velocity field is frictionally screened, and its value at a point is determined by the value of the displacement field (and its gradients) at \emph{that} point. We further demonstrate that an analogous instability afflicts \emph{isotropic} chiral active solids moving in response to an external cue---a situation that is particularly relevant for a wide range of biological systems, including tumours and cellular clusters moving through the extracellular matrix~\cite{rustarazo2026adhesion, spatarelu2019biomechanics,rorth2009collective,ellison2016cell,mugler2016limits,camley2016emergent}. In contrast, in an ideal \textit{uniaxial elastomer}, a higher symmetry---under \emph{independent} rotations of the elastic material and the ordering direction~\cite{Ranjan_Lubensky, Xing_annals, Warner_Terentjev}---completely eliminates this instability, along with one of the two chiral odd elastic moduli that we show characterise three-dimensional odd solids. 

\begin{figure}[t]
    \centering
\includegraphics[width=\linewidth]{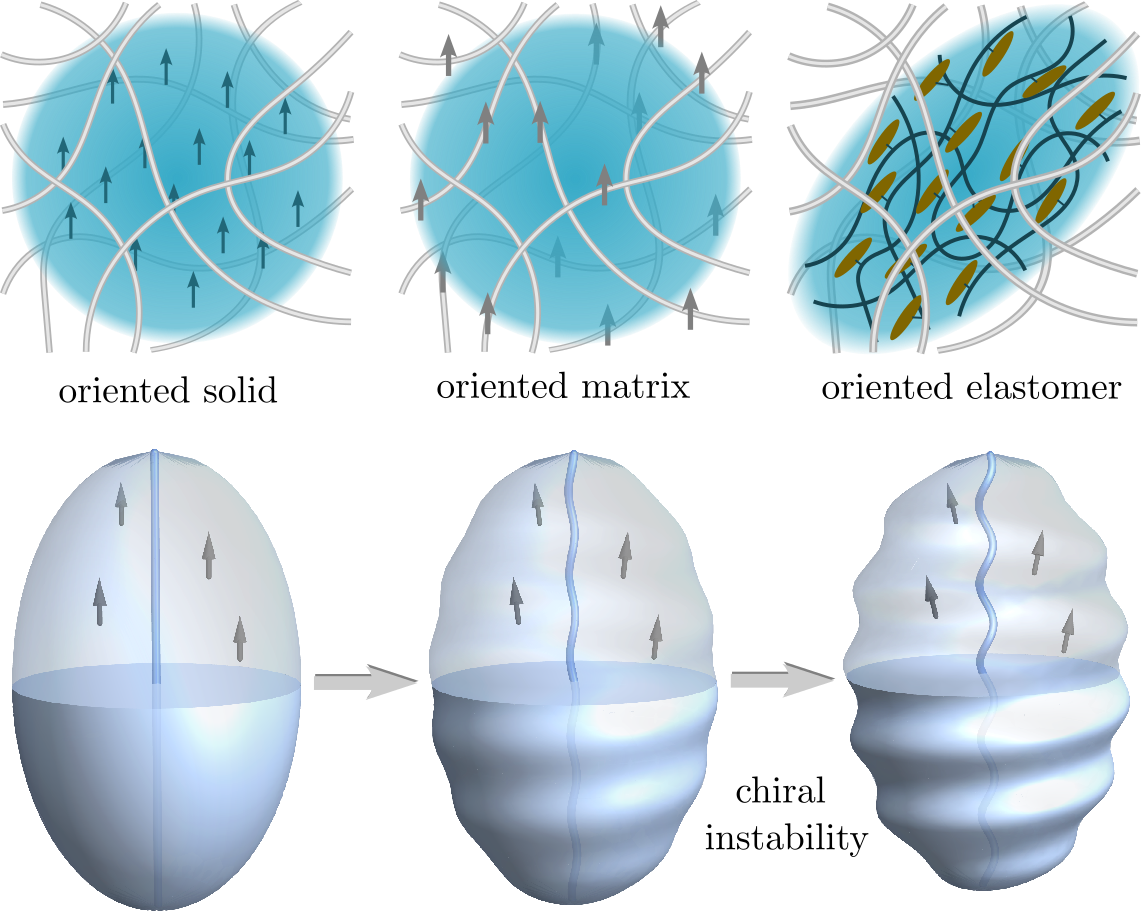}
\caption{Schematic illustration of the types of oriented solids we consider and the generic helical instability in chiral oriented active solids. Top row (left to right): polar solid in a surrounding network; isotropic solid in an polar network; oriented elastomer. Bottom row: Chiral instability. Dark grey arrows indicate the polar axis and we have made the image partly transparent and highlighted a material line parallel to the polar direction to aid visualisation.}
    \label{fig:schematic}
\end{figure}

We start with the linearised theory of the effect of chirality and activity on three-dimensional uniaxial solids moving through a frictional medium. We consider a vacancy-free or permanently crosslinked solid, so that density fluctuations $\delta\rho$ are not an independent dynamical field but are entirely determined by displacement fluctuations ${\bf u}$, measured from an undistorted uniaxial reference state, as $\delta\rho/\rho_0=-\nabla\cdot{\bf u}$, where $\rho_0$ is the mean density. The solid has a \emph{polar} axis along ${\bf n}$, whose mean alignment we take to be ${\bf n}^0\equiv\hat{{\bf z}}$, and therefore moves along that axis in an active system~\cite{AM_sol, Chen_Toner, Mot_sol_PNAS}. The evolution $\dot{\bf u} = \partial_t {\bf u} + {\bf v} \cdot \nabla {\bf u}$ of the displacement field ${\bf u}$ can be written in terms of the velocity ${\bf v}$ as~\cite{ALP} 
\begin{equation}
\label{vstart}
    \dot{u}_i = v_i = v_0 n_i + A_{ijk} {\partial_ju_k} + \mathcal{O}(\nabla^2)\,,
\end{equation}
with self-propulsion at speed $v_0$ along the polar axis \hbox{${\bf n} \simeq [\bsf{I} + (\nabla {\bf u} - \nabla {\bf u}^{\tiny{\rm{T}}})/2]\cdot\hat{\bf z}$}, where $\bsf{I}$ is the identity tensor, to linear order in disturbances, and a strain-dependent active velocity governed by  
\begin{equation} 
\label{generic} 
\begin{split}
&A_{ijk}=a_1 n_i n_j n_k + a_2 n_i(\delta_{jk}-n_jn_k)\\&+a_3(n_j\delta_{ik}+n_k\delta_{ij}-2n_in_jn_k)-a^c(\epsilon_{ilj}n_k+\epsilon_{ilk}n_j)n_l\,, 
\end{split} 
\end{equation} 
whose symmetry in $jk$ ensures rotation invariance of~\eqref{vstart}. The $v_0$ and $a_{1-3}$ terms are \emph{polar}, odd in ${\bf n}$, whereas the only symmetry-allowed chiral active term, with coefficient $a^c$, is invariant under ${\bf n}\to-{\bf n}$, and thus present even in \textit{apolar} uniaxial solids: chiral activity endows even an apolar solid with an $\mathcal{O}(\nabla)$ dynamical response. Elasticity, traditional or odd, is implicitly present at subleading $\mathcal{O}(\nabla^2)$ in~\eqref{vstart}, of a piece with the ordering in achiral polar drifting solids~\cite{AM_sol,Rangan}. The dynamics in the general \textit{polar} case, in a frame co-moving with the mean drift $v_0\hat{{\bf z}}$, reads
\begin{gather}
    \partial_t u_x=a_+\partial_z u_x + a_-\partial_x u_z+{a}^c(\partial_z u_y+\partial_yu_z)+\mathcal{O}(\nabla^2)\,,
    \label{uxad} \\
    \partial_t u_y=a_+\partial_z u_y + a_-\partial_yu_z-{a}^c(\partial_z u_x+\partial_xu_z)+\mathcal{O}(\nabla^2)\,,
    \label{uyad} \\
    \partial_t u_z={a}_1\partial_z u_z+{a}_2\nabla_\perp\cdot{\bf u}_\perp+\mathcal{O}(\nabla^2)\,,
    \label{uzad}
\end{gather}
where $a_{\pm}=a_3 \pm v_0/2$. Considering a perturbation with wavevector ${\bf q}=q_z\hat{{\bf z}}+{\bf q}_\perp$ purely along $\hat{\bf z}$ (i.e., with ${\bf q}_\perp=0$), we see that $u_z$ dynamics decouples from $u_x$ and $u_y$ and has an eigenfrequency $\omega_z=-{a}_1q_z$. The coupled $u_x$ and $u_y$ dynamics have the eigenfrequencies $\omega_\pm=-q_z(a_+ \pm i{a}^c)$. This implies that one of the modes is necessarily unstable at small wavenumber $q\equiv|{\bf q}|$, with zero threshold and irrespective of the sign of ${a}^c$, with a growth rate linear in $q$. Thus activity is guaranteed to destroy a chiral uniaxial solid, whether polar or apolar, in an inert background medium. Elastic force densities appear at $\mathcal{O}(q^2)$ and thus cannot prevent this instability at small-enough wavenumbers, but do cut it off at larger wavenumbers, with the fastest growing mode being given by the ratio of $a^c$ and the elastic coefficient penalising $\partial_z^2{\bf u}_\perp$ fluctuations (assuming that the latter is positive), where ${\bf u}_\perp\equiv (u_x,u_y)$. 
The unstable mode is a helical transverse displacement field, right- or left-handed according as $a^c>0$ or $<0$, and propagating along $\hat{\bf z}$ with speed $-a_+$ in a polar material. This is illustrated in Fig.~\ref{fig:schematic}. The helical unstable mode structure is the same as that suffered by the director field in passive (and active) chiral uniaxial fluids, which in that case results in the formation of usual cholesteric states. As remarked at the beginning of this Letter, a possible final state of the present instability is a solid with a helical orientation field, self-rotating because of activity. This generic instability, distinct from the (non-generic) instability of chiral active filaments in~\cite{RJoy}, is a key result of this article. 

We now argue that an analogous generic instability affects an idealized \emph{isotropic} chiral elastic continuum drifting in an externally determined direction. This case is potentially relevant for chiral soft-matter systems moving, for instance, due to an external field~\cite{drifting_skyrmions_1,drifting_skyrmions_2,Skyrmion_schools}, as well as for tissues or cell clusters moving in response to an external cue~\cite{spatarelu2019biomechanics,rorth2009collective,ellison2016cell,mugler2016limits,camley2016emergent}. The chiral force densities still enter through a term $A_{ijk}\partial_j u_k$ but with a distinguished direction $\hat{{\bf z}}$; $A_{ijk}$ is symmetric under interchange of $j$ and $k$ only if both $\in\{x,y\}$. This allows for chiral force densities $a^c_1\epsilon_{ijz}\partial_zu_j+a^c_2\epsilon_{ijz}\partial_j{ u}_z$. For a perturbation with wavevector along $\hat{{\bf z}}$, to leading order in wavenumbers, this implies an eigenfrequency with an imaginary part $\mathrm{Im}[\omega_\pm]=\pm a_1^cq_z$ which demonstrates that the solid state is generically unstable irrespective of the sign of $a_1^c$. 

Note that such a solid, crystalline or amorphous, is destabilised even when there is no axis about which rotation leaves the system invariant. For instance, for an elastic continuum in a medium with cubic anisotropy, an active force density $\propto a_3^c\nabla\times{\bf u}$ is present. A displacement field dynamics containing only this term at first order in gradients has a pair of eigenfrequencies at $\mathcal{O}(q)$ (with the third vanishing at this order in $q$): $\omega_\pm=\pm i a_3^c q$ again implying a generic instability. This eigenfrequency does not depend on the \emph{direction} of the wavevector, implying that a perturbation along any wavevector direction is unstable. 

We now show that a higher-symmetry uniaxial solid state---a uniaxial elastomer---escapes the generic instability discussed above; in fact, chiral activity does not affect this material at $\mathcal{O}(\nabla)$. Uniaxial elastomers are invariant under \emph{independent} rotations of the ordering direction and the solid structure in contrast to traditional uniaxial solids we have discussed till now which are only invariant under \emph{joint} rotations of the asymmetry axis and the structure. While we could directly construct the displacement dynamics of the active and chiral elastomeric state (as we did for uniaxial and chiral active solids) by implementing this additional invariance, we follow the more straightforward and pedagogically clearer path of obtaining the dynamics of the uniaxial phase by introducing spontaneously broken rotation symmetry into the theory of a reference \emph{isotropic} elastomer. 

We define a linearised strain field ${\bsf U}={\bsf C}+(1/3)\Phi{\bsf I}$, $U_{ij}=(\partial_j\bar{u}_i+\partial_i\bar{u}_j)/2$, with $\bar{{\bf u}}$ being a displacement field about an \emph{isotropic} state and ${\bsf C}$ and $\Phi$ being the trace-removed and isotropic parts of ${\bsf U}$. We further define a traceless symmetric nematic order parameter ${\bsf Q}$ which, in a polar state with a vectorial order parameter ${\bf p}\neq 0$, is well approximated by ${\bf pp}-(1/3){\bsf I}p^2$, but is also nonzero in an apolar but uniaxial state in which ${\bf p}=0$. The velocity field in terms of these variables is 
    \begin{equation}
        v_i = w_1(\Phi)p_i+w_2C_{ij}p_j+w^c_1\epsilon_{ikl}Q_{lm}C_{mk}\,,
    	\label{frc}
    \end{equation}
where we have only retained terms that finally contribute to the displacement field equation in the elastomeric state, to leading order in gradients, upon rotation symmetry being spontaneously broken. We will examine the effect of polar and chiral activity on an equilibrium uniaxial elastomeric state. For this, we follow~\cite{Ranjan_Lubensky, AM_sol} and use the free energy
\begin{equation}
    F=\int_{\bf x} \left [ \frac{\lambda}{2}\left(\Phi-\frac{s}{\lambda}p^2\right)^2+\mu\text{Tr}\left( {\bsf C}-\frac{t}{\mu}{\bsf Q}\right)^2+f_p \right ] \,,
    \label{FQ}
\end{equation}
with $f_p=-({r}/{2})p^2+({w}/{4})p^4+(K/2)(\nabla{\bf p})^2$. To this order in gradients and fields $F$ has no contribution due to chirality. The generic free-energy couplings $s$ and $t$ between the strain and the polarisation lead to the development of a strain anisotropy either parallel ($t>0$) or perpendicular ($t<0$) to ${\bsf Q}$. 

A polar uniaxial state is realised when $r>0$ with ${\bf p}=p_0 \hat{\bf z}$, with $p_0=\sqrt{r/w}$, and a mean strain whose isotropic and deviatoric parts are $\Phi_0=sp_0^2/\lambda$ and ${\bsf C}_0=(t/\mu){\bsf Q}_0$, respectively. We will examine fluctuations about this state which, from Eq.~\eqref{frc}, implies a steady-state velocity ${\bf v}_0=[p_0w_1(\Phi_0)+2tw_2p_0^3/3\mu]\hat{\bf z}$. Of course, chiral activity does not affect the steady-state speed: ${\bsf C}_0\propto{\bsf Q}_0$ and therefore, $\epsilon_{ikl}Q_{0_{lm}}C_{0_{mk}}$ vanishes.

From Eq.~\eqref{FQ} we see that in the uniaxially ordered state, orientational fluctuations are determined by strain fluctuations: $\delta{\bf p}_\perp=(\mu/tp_0)\delta{\bsf C}_{z\perp}$ \cite{Ranjan_Lubensky} and $\delta p_z=(6t\delta C_{zz}+3s\delta\Phi)/\bar{r}p_0$, where $\bar{r}=3w+6s^2/\lambda+8t^2/\mu$. This observation is enough to demonstrate that chiral activity does not destabilise \textit{elastomers}. Expanding $\epsilon_{ikl}Q_{lm}C_{mk}$ to linear order in fluctuations and using the result discussed above for fluctuations of $\delta{\bf p}$, we find that it vanishes to this order. To see why, notice that ${\bsf Q}^0$ and ${\bsf C}^0$ are proportional so that, to linear order, $\epsilon_{ikl}Q_{lm}C_{mk} \approx \epsilon_{ikl} \delta Q_{lm}C^0_{mk}+\epsilon_{ikl}Q^0_{lm}\delta C_{mk}$. Since the $\delta{\bsf Q}$ fluctuations relax to $\delta{\bsf C}$ ones, the tensor fully contracted with $\bm{\epsilon}$ is symmetric, and therefore, this term vanishes. We again emphasise that this vanishing of the chiral active force density in the elastomeric state is due to the invariance of polar elastomers under independent rotations in both reference and target spaces~\cite{Ranjan_Lubensky, Olmsted}. This is in contrast with polar or nematic liquid crystals, which are invariant only under \emph{joint} rotations of the anisotropy direction and the coordinates, and uniaxial solids, in which the symmetry is under joint rotations of the elastic material and the anisotropy direction. 

Thus, to $\mathcal{O}(\nabla)$, chirality does not affect active polar elastomers. The dynamics is best presented in terms of a displacement field ${\bf u}$ defined about the uniaxial state (which has a strain ${\bsf C}_0$ with respect to an isotropic reference). The straightforward transformation from $\bar{{\bf u}}$ to ${\bf u}$ was discussed in detail in~\cite{Ranjan_Lubensky, AM_sol} and is briefly summarised in the accompanying long paper~\cite{ALP}. The dynamical equations for an active and chiral polar elastomer to $\mathcal{O}(\nabla)$---which are the same as that of polar active elastomers---was discussed in~\cite{AM_sol, AM_sol_err}; for completeness, we also present them in the accompanying long paper~\cite{ALP}.

This high-symmetry elastomeric state is at best approximately realised in physical systems: as discussed in~\cite{Ranjan_Lubensky, Warner_Terentjev}, the memory of the crosslinking direction persists in elastomeric materials due to disorder, breaking the additional symmetry under independent rotation of the anisotropy direction and the elastomer to one under only joint rotation. In this case, it has the same symmetries as a uniaxial or polar solid discussed earlier. Indeed, we show in~\cite{ALP} that the chiral active force density does not vanish in this case and destabilises the physical elastomeric state.

We now turn to a discussion of the effect of chiral activity on three-dimensional uniaxial solids and elastomers when momentum is conserved. In this case, all force densities \emph{must} arise from the divergence of a stress tensor which we take to be symmetric for conservation of angular momentum. Since $\bm{\sigma}$ must be constructed from gradients of the displacement field, all such force densities arise at least at $\mathcal{O}(\nabla^2)$ and none of the ${\cal O}(\nabla)$ effects discussed up to this point survives in a momentum-conserving system. Instead, the physics is controlled by elasticity, both conventional and odd.

We first discuss the form of $\bm{\sigma}$ for a chiral and active uniaxial solid and subsequently comment on the differences in an elastomer. On symmetry grounds a polar solid has eight elastic moduli of which three are odd. The conventional stresses are discussed in detail elsewhere e.g.,~\cite{Ranjan_Lubensky}; we record only that they are all invariant under inversion of the polar direction, ${\bf n} \to - {\bf n}$. We focus now on the odd contributions. Defining the symmetrised strain tensor $E^{S}_{ij} = 1/2 [\partial_i u_j + \partial_j u_i]$, for displacements about the polar elastomeric state, the odd elastic stress has the form 
\begin{widetext}
\begin{equation}
  {\sigma^{o}_{ij} = 2 \left[\epsilon_{ikm}n_m \left\{ 2\zeta^c_1 \,n_jn_l + \zeta^c_2 \left( \delta_{jl} -n_jn_l \right) \right\} E^{ST}_{kl} \right]^S + \zeta^{on} \bigl( n_i n_j \delta_{kl} - \delta_{ij} n_k n_l \bigr) E^{S}_{kl}} \,,
\end{equation}
\end{widetext}
and implies an odd elastic force density 
\begin{equation}
\begin{split}
    & \nabla\cdot\bm{\sigma}^{o} = \zeta^c_{1} \,\partial_z \bigl( 2E^A_{xy} \,\hat{\bf z} + \partial_z \bm{\varepsilon}\cdot{\bf u}_{\perp} + \bm{\varepsilon}\cdot\nabla_{\perp} u_z \bigr) \\
    & \quad + \zeta^c_2 \,\nabla_\perp^2 \bm{\varepsilon} \cdot{\bf u}_\perp + \zeta^{on} \bigl( \partial_z (\nabla_{\perp} \cdot {\bf u}_{\perp}) \,\hat{\bf z} - \nabla_{\perp} \partial_z u_z \bigr)  \,,
\end{split} 
\label{eq:odd_force}
\end{equation}
where $E^A_{ij}=1/2(\partial_iu_j-\partial_ju_i)$ is the antisymmetric part of the displacement gradient tensor and $\varepsilon_{ij} \equiv \epsilon_{ijz}$ is the \emph{two-dimensional} Levi-Civita tensor. 
The contributions with coefficients $\zeta^c_1$ and $\zeta^c_2$ are chiral and polar; that with coefficient $\zeta^{on}$ is nonchiral and allowed in uniaxial materials. 

The odd elastic force density~\eqref{eq:odd_force} is an elastic analogue of the three-dimensional odd viscous force density discussed in~\cite{khain2022stokes, SJ_col}. The force density with the coefficient $\zeta^c_2$ also appears in two-dimensional solids~\cite{scheibner2020odd, AM_hex, AM_chi_rev, fruchart2023odd} and three-dimensional columnar materials~\cite{SJ_col}, however, the one with the coefficient $\zeta^c_1$ has not been discussed previously. In the Stokesian regime these chiral odd elasticities lead to a pair of eigenfrequencies with a nonvanishing \emph{real} (i.e. oscillatory) part at zero wavenumber. For a wavevector purely along $\hat{{\bf z}}$, $\lim_{q_z\to 0}\mathrm{Re}[\omega_\pm]=\pm \zeta^c_{1} /\eta$, where $\eta$ is an isotropic viscosity (we have neglected odd viscosity here for clarity but present the detailed dynamics including odd viscosity in the accompanying long paper~\cite{ALP}). The third odd elastic stress ($\zeta^{on}$) has no viscous analogue for incompressible uniaxial fluids. It allows the coefficients of the terms $\partial_z( \nabla_\perp\cdot{\bf u}_\perp)\hat{{\bf z}}$ and $\nabla_\perp\partial_zu_z$---equal in equilibrium systems---to be distinct in active materials~\cite{ALP, AM_sol}. Such a stress~\cite{zahalak1996non} was invoked to account for the dynamics of sarcomeres in~\cite{shankar2024active} and was implicitly obtained in~\cite{AM_sol}, though that work did not attribute this contribution to odd elasticity or recognise that it violates the Maxwell-Betti relation. 

The odd elastic force density~\eqref{eq:odd_force} depends on both the wavevector ${\bf q}$ of a displacement mode and on the elastic moduli, allowing for significant variation in behaviour for materials with different moduli. The coefficient of the two-dimensional term ${\bm \varepsilon} \cdot {\bf u}_{\perp}$ is $\zeta^c_1 q_z^2 + \zeta^c_2 q_{\perp}^2$ and has a consistent sign for all wavevectors if $\zeta^c_1, \zeta^c_2$ have the same sign, but is variable if they have opposite signs. These correspond to topologically distinct behaviours for three-dimensional odd elastic force densities~\cite{OddTopology}. 

We now demonstrate that the higher symmetry of elastomers has an important consequence on odd elastic force densities. As earlier, we develop the theory starting from a coupled description involving a polar order parameter and a strain tensor about an isotropic reference. The generic active stress, containing both achiral and chiral parts, can be taken as 
\begin{widetext}
    \begin{equation}
        \sigma^{a}_{ij} = \zeta_1 Q_{ij} + 2\left[ \epsilon_{ilk}p_k \left\{ 2 \zeta^c_1 p_jp_m + \zeta^c_2 \left( p^2 \delta_{jm} - p_jp_m \right) \right\} C_{lm} \right]^S\,{,}
    	\label{strs}
    \end{equation}
\end{widetext}
where, we remind the reader, ${\bsf C}$ is the strain tensor with respect to the isotropic state. We do not include an explicit $\zeta^{on}$ term as one is generated in expansion about the uniaxial state. The consequence of the first achiral and apolar term in~\eqref{strs} was examined in~\cite{AM_sol}. Briefly, the enhanced symmetry of an elastomer allows for only four elastic force densities in equilibrium systems, instead of the five in uniaxial solids, with no force densities $\propto\nabla_\perp^2 u_z\hat{{\bf z}}$ or $\partial_z^2{\bf u}_\perp$ (in terms of displacement variables defined with respect to the uniaxial reference state); the active term $\zeta_1$ yields force densities that are $\propto\nabla_\perp^2 u_z\hat{{\bf z}}$ or $\partial_z^2{\bf u}_\perp$ which, when stabilising, makes the linearised dynamics of achiral elastomers equivalent to that of uniaxial solids. $\zeta_1$ also naturally generates the achiral odd elasticity ($\zeta^{on}$) discussed above: as shown in~\cite{AM_sol} and~\cite{ALP}, the coefficients of $\partial_z\nabla_\perp\cdot{\bf u}_\perp\hat{{\bf z}}$ and $\nabla_\perp\partial_zu_z$ differ by a $\zeta_1$-dependent term. We now focus on the explicitly chiral terms. The $\zeta^c_1$ stress can be written $4 \zeta^c_1 [ ( \epsilon_{ikl} p_l p_m C_{mk} ) p_j ]^S$ and the term in round brackets is then the same as the chiral term $\propto \epsilon_{ikl} Q_{lm} C_{mk}$ we analysed previously in the frictional dynamics and showed vanished in an elastomer. Thus, in contrast to a uniaxial solid, the \emph{only} chiral odd elastic force density in a uniaxial elastomer is $\propto \zeta^c_2 \nabla_\perp^2 \bm{\varepsilon} \cdot{\bf u}_\perp$, i.e., the one also present in two-dimensional odd solids. This is another consequence of the elastomeric symmetry under independent rotation of the polar direction and the solid structure.

That there is only one chiral active elasticity in polar elastomers, while a straightforward consequence of symmetry, merits discussion. In equilibrium uniaxial elastomers, $u_z$ is equivalent to a smectic displacement field, whereas ${\bf u}_\perp$ is equivalent to a columnar displacement field. Reference~\cite{AM_sol} showed that this equivalence carries over to achiral active systems: activity endows elastomers with force densities that are equivalent to introducing a tilt modulus in a smectic (i.e., $\nabla_\perp^2 u_z\hat{{\bf z}}$ \cite{Tapan}) and a column tension in a columnar phase (i.e., $\partial_z^2{\bf u}_\perp$ \cite{SJ_col}). We show here that this mapping continues to hold even in chiral systems. Active smectics have no chirality-induced elastic force density at $\mathcal{O}(\nabla^2)$ \cite{SJ_smec}, polar columnar systems have one \cite{SJ_col} $\propto\nabla_\perp^2 \bm{\varepsilon} \cdot{\bf u}_\perp$, as here. We display detailed dynamical equations for momentum-conserving chiral and polar active elastomers in the accompanying long paper~\cite{ALP}.

In this Letter, we have constructed a comprehensive theory of chiral activity in three-dimensional solids and elastomers. We have uncovered a generic instability due to chiral activity that destroys the solid state in a fixed background medium. Importantly, this is not due to traditional elasticity but a force density that arises at a lower order in gradients. We have shown that elastomers escape this instability due to their enhanced symmetry. We further show that while momentum-conserving polar solids have two odd elastic force densities due to chiral activity, elastomers---their higher-symmetry cousins---have only one. Our study, which is the first to look at the effect of chirality in three-dimensional active solids, should be of relevance for experiments on a variety of biological systems, including for the transport of cellular clusters or tissues through the extracellular matrix either in response to an external cue~\cite{spatarelu2019biomechanics,rorth2009collective,ellison2016cell,mugler2016limits,camley2016emergent} or by spontaneously breaking rotation symmetry~\cite{AM_tissue1, AM_tissue2,piotrowski2017dynamics}. Our results should also apply to driven skyrmion lattices~\cite{drifting_skyrmions_1,drifting_skyrmions_2,Skyrmion_schools}. 

\begin{acknowledgments} 
	AM acknowledges a TALENT fellowship awarded by CY Cergy Paris Universit\'e and the support of ANR through the grant PSAM. AM, SJK and SR thank the Isaac Newton Institute for Mathematical Sciences, Cambridge, for support and hospitality during the programmes ``New Statistical Physics in Living Matter: non-equilibrium states under Adaptive Control'' and ``Anti-diffusive dynamics: from sub-cellular to astrophysical scales'' (EPSRC grant no. EP/R014604/1), and SR additionally for a Rothschild Distinguished Visiting Fellowship. This research was supported in part by grant NSF PHY-2309135 to the Kavli Institute for Theoretical Physics. SR was supported in part by a J C Bose Fellowship of the ANRF, India. SJK was supported by the INI-Simons research fellowship, Cambridge.
\end{acknowledgments}

\bibliography{ref}

\end{document}